\documentclass[aps,prb,twocolumn,showpacs,preprintnumbers]{revtex4}
\usepackage{graphicx} %Include figure files,superscriptaddress
\usepackage{amsmath}
\usepackage{graphicx,epstopdf}
\begin{document}

\title{Double phase transition in the triangular antiferromagnet Ba$_3$CoTa$_2$O$_9$}
\author{K. M. Ranjith}
\affiliation{School of Physics, Indian Institute of Science Education and Research Thiruvananthapuram-695016, India}
\author{K. Brinda}
\affiliation{School of Physics, Indian Institute of Science Education and Research Thiruvananthapuram-695016, India}
\author{U. Arjun}
\affiliation{School of Physics, Indian Institute of Science Education and Research Thiruvananthapuram-695016, India}
\author{N. G. Hegde}
\affiliation{School of Physics, Indian Institute of Science Education and Research Thiruvananthapuram-695016, India}
%\author{A. A. Tsirlin}
%\affiliation{Experimental  Physics  VI,  Center  for  Electronic  Correlations  and  Magnetism, University  of  Augsburg,  86159  Augsburg,  Germany}
\author{R. Nath}
\email{rnath@iisertvm.ac.in}
\affiliation{School of Physics, Indian Institute of Science Education and Research Thiruvananthapuram-695016, India}
\date{\today}

\begin{abstract}
Here, we report the synthesis and magnetic properties of a new triangular lattice antiferromagnet Ba$_3$CoTa$_2$O$_9$. The effective spin of Co$^{2+}$ is found to be $J=1/2$ at low temperatures due to the combined effect of crystal field and spin-orbit coupling. Ba$_3$CoTa$_2$O$_9$ undergoes two successive magnetic phase transitions at $T_{N1}\simeq0.70$~K and $T_{N2}\simeq0.57$~K in zero applied field, which is typical for triangular antiferromagnets with the easy-axis magnetic anisotropy. With increasing field, the transition anomalies are found to shift toward low temperatures, confirming the antiferromagnetic nature of the transitions. At higher fields, the transition peaks in the heat capacity data disappear and give way to a broad maximum, which can be ascribed to a Schottky anomaly due to the Zeeman splitting of spin levels. The $H-T$ phase diagram of the compound shows three distinct phases. The possible nature of these phases is discussed.
\end{abstract}

\maketitle

\section{Introduction}
Magnetic frustration and the rich variety of phases driven by competing magnetic couplings have attracted a lot of attention in present-day condensed matter physics.\cite{Mila2011} Two dimensional (2D) triangular lattice antiferromagnet (TLAF) is the simplest example of a geometrically frustrated quantum system, where lattice geometry precludes simultaneous minimization of exchange interaction energy on different bonds, thus leading to a highly degenerate classical ground state.\cite{Coldea2001,Coldea2003,Dong2000} Quantum fluctuations, which are most pronounced in systems with reduced dimensionality and low spin values, lift the classical degeneracy and stabilize a variety of exotic phases, including quantum spin liquid (QSL),\cite{Gardner1999,Yamashita2008,Balents2010,Yamashita2010} spin ice,\cite{Bramwell2001} and field-induced states manifesting themselves by plateau features in the magnetization.\cite{Misguich2001,Kodama2002} In a quasi-2D isotropic Heisenberg TLAF, the spins order antiferromagnetically in a 120$^{\circ}$ structure at zero field. Under external magnetic field this 120$^{\circ}$ ordered state evolves to an `up-up-down' ($uud$) state showing a plateau at the 1/3 of the saturation magnetization, through quantum and/or thermal fluctuations, as in the $S=1/2$ compounds Cs$_2$CuBr$_4$ and Ba$_3$CoSb$_2$O$_9$.\cite{Ono2003,Fortune2009,Susuki2013} At very high fields, the $uud$ state becomes unstable leading to canted spin states. Ground states of TLAFs are also sensitive to inter-layer coupling and exchange anisotropy which often lead to even more complex phases.\cite{Alicea2009,Coldea2003}

Recently, a family of TLAFs, Ba$_3MM'_2$O$_9$ ($M=$ Co, Ni, Cu, Mn and $M'=$ Sb and Nb), has been studied extensively. These studies unveiled a plethora of interesting properties.\cite{Susuki2013,Zhou2012,Yokota2014,Shirata2011,Koutroulakis2015,Zhou2011,Hwang2012,Doi2004,Lee2014,Lee2014b} The $S=1/2$ compound Ba$_3$CuSb$_2$O$_9$ shows features of a QSL, such as the absence of magnetic long-range ordering (LRO) down to 0.2~K despite the large Curie-Weiss temperature $\theta_{\rm CW}$, and the linear temperature dependence of the heat capacity at low temperatures.\cite{Zhou2011} Owing to the octahedral crystal field and the spin-orbit coupling, the Co$^{2+}$ ion in Ba$_3$CoSb$_2$O$_9$ and Ba$_3$CoNb$_2$O$_9$ features an effective spin $S=1/2$ at low temperatures. While Ba$_3$CoNb$_2$O$_9$ shows two consecutive phase transitions, Ba$_3$CoSb$_2$O$_9$ has only one transition in zero field, likely due to the easy-axis and easy-plane type anisotropies, respectively.\cite{Lee2014,Zhou2012} Ba$_3$CoSb$_2$O$_9$ is reported to display magnetization plateaus at intermediate fields below the saturation field.\cite{Susuki2013} Because of such non-trivial and exotic properties, Co$^{2+}$ based TLAFs are paid a great deal of attention both experimentally and theoretically.

\begin{figure}
\begin{center}
\includegraphics[scale=0.8]{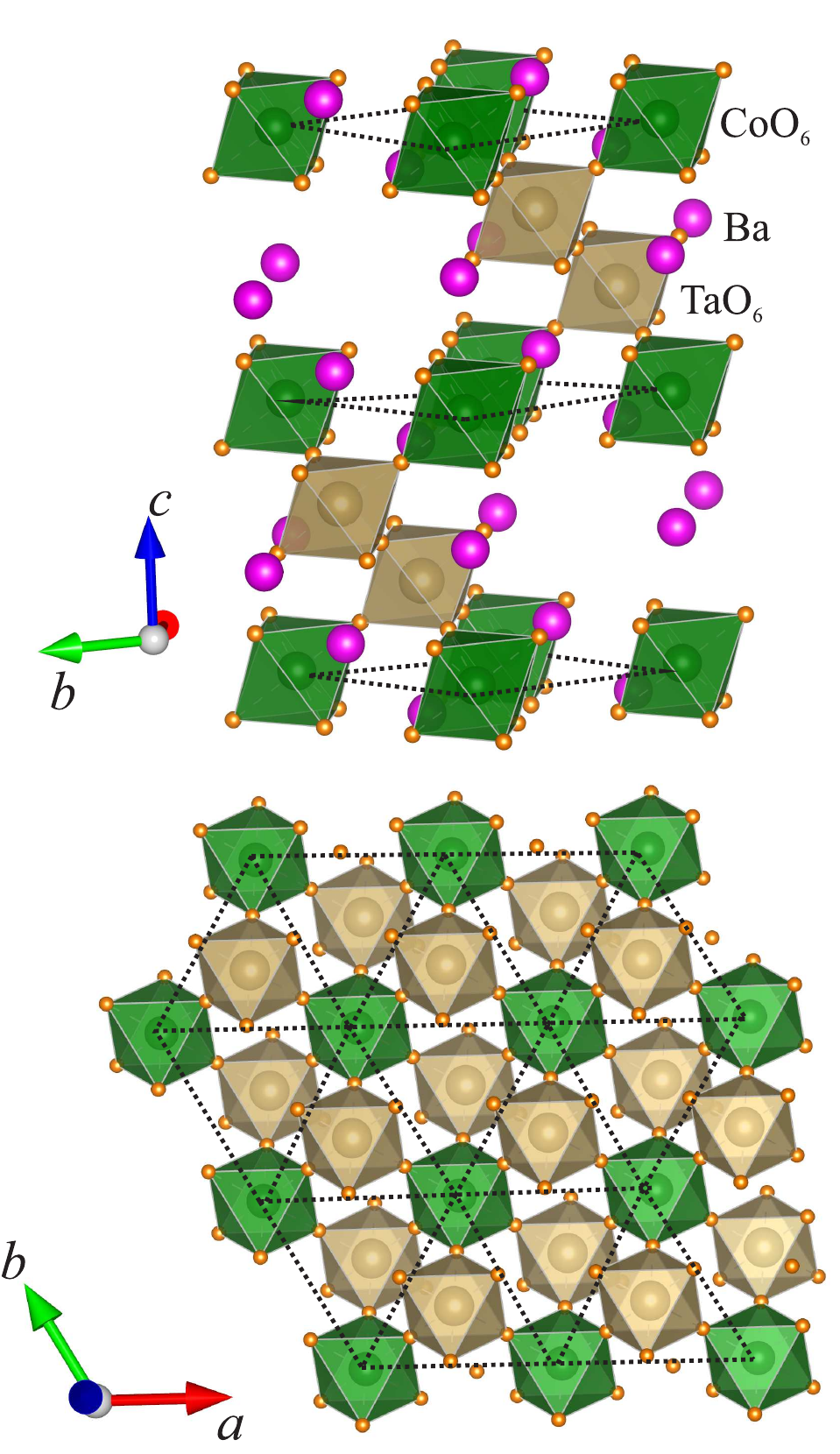}
\caption{Upper panel: Schematic crystal structure of Ba$_3$CoTa$_2$O$_9$. Lower panel: Triangular layer of Co$^{2+}$ ions formed by the corner sharing of CoO$_6$ and TaO$_6$ octahedra in the $ab-$plane.}
\label{structure}
\end{center}
\end{figure}
In this paper, we report the magnetic behavior of a new compound of this family, Ba$_3$CoTa$_2$O$_9$. It crystallizes in a hexagonal structure \cite{Treiber1982} with the space group $P\bar{3}m1$. Figure~\ref{structure} shows the crystal structure of Ba$_3$CoTa$_2$O$_9$, which can be represented as a framework consisting of CoO$_6$ octahedra sharing corners with dimers of TaO$_6$ octahedra. The Co$^{2+}$ ions, which occupy the $1b$ site, form triangular lattices parallel to the $ab-$plane (see the bottom panel of Fig.~\ref{structure}) and are separated by non-magnetic Ba atoms. A weak inter-layer coupling still can be envisaged via an extended Co$^{2+}$-O$^{2-}$-Ta$^{5+}$-O$^{2-}$-Co$^{2+}$ pathway. Our magnetic measurements reveal two magnetic phase transitions at $T_{N1}\simeq0.70~$K and $T_{N2}\simeq0.57~$K with a complex $H-T$ phase diagram.

%The Co$^{2+}$-Co$^{2+}$ distances ($\sim 5.77$~\AA) along the edges of the triangle and the bond angles $\angle$ Co$^{2+}$-Ta$^{5+}$-Co$^{2+}$ between them are same ($\sim 89.96^{\circ}$) implying a nearly isotropic triangular lattice in the $ab$-plane. Moreover, the CoO$_6$ octahedron is also isotropic with Co-O distances ($\sim 1.55$~\AA) same along all bonds. Because of the highly symmetric structure, Dzyaloshinsky-Moriya (DM) interaction which is one of the leading asymmetric term in the Hamiltonian is absent here. The absence of DM interaction makes the exchange mechanism rather simpler and easier to understand the ground states, as compared to the spatially distorted compound, Ba$_3$CoSb$_2$O$_9$ of this family.\cite{Starykh2014}

\section{Experimental details}
Polycrystalline sample of Ba$_3$CoTa$_2$O$_9$ was prepared by a conventional solid-state reaction technique using BaCO$_3$ (99.999\%, Aldrich), CoO (99.99\%, Aldrich), and Ta$_2$O$_5$ (99.99\%, Aldrich) as starting materials. Stoichiometric mixture of the starting materials was intimately ground, pressed into pellets, and fired at 900$~^\circ$C for 8~hours and then at 1200$~^\circ$C for 48 hours in air with intermediate grindings and pelletizations. Phase purity of the sample was confirmed by recording powder x-ray diffraction (XRD) pattern using the PANalytical powder diffractometer (CuK$_\alpha$ radiation, $\lambda_{\rm ave}$ = 1.5406~\AA) at room temperature. Le-bail fit of the observed XRD pattern was performed using \verb"FULLPROF" package \cite{Carvajal1993} for which the initial parameters were taken from Ref.~\onlinecite{Treiber1982}.
\begin{figure}
\begin{center}
\includegraphics{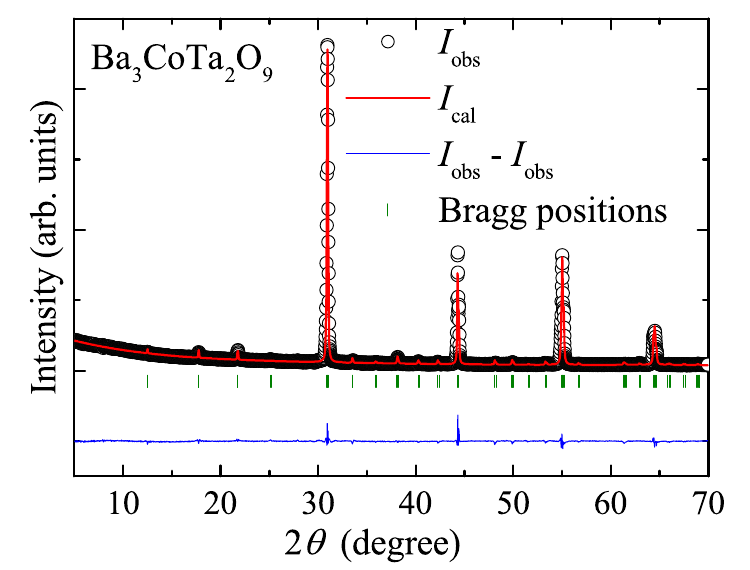}
\caption{\label{refn} Powder XRD data for Ba$_3$CoTa$_2$O$_9$ collected at room temperature. The solid line represents the Le-bail fit of the data. The Bragg peak positions are indicated by green vertical bars, and the bottom solid blue line indicates the difference between the experimental and calculated intensities.}
\end{center}
\end{figure}
Figure~\ref{refn} shows the room temperature powder XRD pattern for Ba$_3$CoTa$_2$O$_9$ along with its Le-bail fit. All the peaks in the XRD pattern could be indexed based on the space group $P\bar{3}m1$ (No. 164). The lattice parameters obtained from the Le-bail fit [$a = b = 5.7748(6)$~\AA~and $c = 7.0908(1)$~\AA] are comparable to the previous report [$a = b = 5.7750 $~\AA~and $c= 7.0960$~\AA].\cite{Treiber1982} The goodness of fit parameter was obtained to be $\chi^2\simeq4.58$.

Magnetization ($M$) measurements were performed as a function of temperature $T$ and applied field $H$ using vibrating sample magnetometer (VSM) attachment to the physical property measurement system (PPMS, Quantum Design). Heat capacity $C_{\rm p}$ as a function of $T$ and $H$ was measured on a pressed pellet using the relaxation technique in a PPMS. The low temperature ($T \leq 2$~K) $C_{\rm p}$ measurements were carried out using an additional $^{3}$He attachment.

\section{Results and Discussion}
\subsection{Magnetization}
\begin{figure}
\begin{center}
\includegraphics{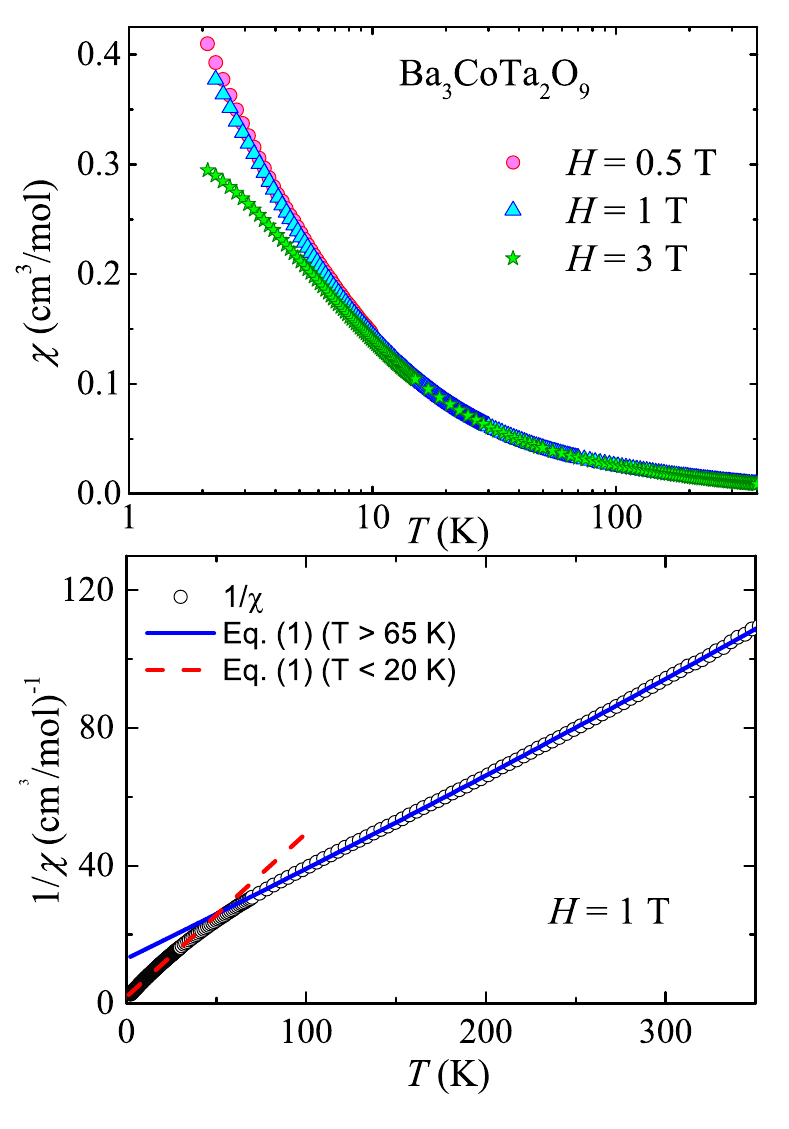}
\caption{Upper panel: $\chi(T)$ of Ba$_3$CoTa$_2$O$_9$ measured at different applied magnetic fields. Lower panel: Inverse magnetic susceptibility (1/$\chi$) of Ba$_3$CoTa$_2$O$_9$ as a function of temperature measured at 1~T along with the fits by Eq.~\eqref{cw}.}
\label{Chi}
\end{center}
\end{figure}
Temperature-dependent magnetic susceptibility $\chi(T)$~$(\equiv M/H)$ measured at applied fields of $H=0.5$~T, 1~T, and 3~T are shown in the upper panel of Fig.~\ref{Chi}. With decreasing $T$, $\chi(T)$ increases in a Curie-Weiss (CW) manner as expected in the paramagnetic regime. No indication of a magnetic LRO was observed down to 2~K. As shown in the lower panel of Fig.~\ref{Chi}, $1/\chi$ varies linearly with $T$ in the high-$T$ regime and a change of slope is observed at low temperatures. In order to extract the magnetic parameters, $\chi(T)$ measured at $H=1~$T in the high-$T$ regime was fitted by the following expression:
\begin{equation}\label{cw}
\chi(T) = \chi_0 + \frac{C}{(T + \theta_{\rm CW})},
\end{equation}
where $\chi_0$ is the temperature-independent contribution consisting of core diamagnetism of the core electron shells ($\chi_{\rm core}$) and Van-Vleck paramagnetism ($\chi_{\rm VV}$) of the open shells of the Co$^{2+}$ ions present in the sample. The second term in Eq.~\eqref{cw} is the CW law with the CW temperature ($\theta_{\rm CW}$) and Curie constant $C=N_{\rm A}\mu_{\rm eff}^2/3k_{\rm B}$, where $N_{\rm A}$ is Avogadro's number, $k_{\rm B}$ is Boltzmann constant, $\mu_{\rm eff}=g\sqrt{S(S+1)}\mu_{\rm B}$ is the effective magnetic moment, $g$ is the Land$\acute{\rm e}$ $g$-factor, and $S$ is the spin quantum number.

Our CW fit in the high-temperature range ($T > 65$~K) yields $\chi_0\simeq-6.91 \times 10^{-4}$~cm$^3$/mol, $C\simeq3.96$~cm$^3$K/mol, and $\theta_{\rm CW}\simeq50.1$~K. From this value of $C$, the effective moment was calculated to be $\mu_{\rm eff}~[=\sqrt{3k_{\rm B}C/N_{\rm A}}] \simeq 5.63~\mu_{\rm B}$/Co. This value of $\mu_{\rm eff}$ is close to the value of $\sim 6~\mu_{\rm B}$ [$=g\sqrt{S(S+1)}\mu_{\rm B}$] expected for the high-spin state ($S=3/2$) of Co$^{2+}$ taking $g\simeq3.1$, obtained from the magnetization data (discussed later).

Since a change in slope was observed at low temperatures for $1/\chi$, the data were fitted separately using Eq.~\eqref{cw} for $T < 20$~K, which yields $\chi_0\simeq0.0355$~cm$^3$/mol, $C\simeq1.2$~cm$^3$K/mol, and $\theta_{\rm CW}\simeq 1$~K. From this value of $C$, the effective moment was calculated to be $\mu_{\rm eff} \simeq 3.1~\mu_{\rm B}$/Co, which is reminiscent of an effective $S=\frac12$ state with the same $g\simeq3.1$ (the value of $\mu_{\rm eff}=2.7~\mu_{\rm B}$ would be expected). The reduction in $\mu_{\rm eff}$ upon cooling is indeed expected for Co$^{2+}$ in the octahedral environment, because spin-orbit coupling $\lambda$ splits the lowest orbital triplet into six Kramers doublets. When the temperature is low enough such that $T\ll \lambda/k_B$, magnetic behavior is determined by the lowest Kramers doublet with $J=\frac12$, where $J$ stands for the full angular momentum, as opposed to the spin angular momentum $S$. In Co$^{2+}$ compounds, $|\lambda|/k_B\simeq250~$K \cite{Shirata2012} and hence at low temperatures Ba$_3$CoTa$_2$O$_9$ is expected to produce an effective spin-1/2 behavior. The effective spin-$1/2$ ground state is also reported for other TLAFs with the octahedral coordinates Co$^{2+}$ sites, such as Ba$_3$Co$M_2$O$_9$ ($M$~=~Sb, Nb) \cite{Doi2004,Lee2014} and $A$Co$B_3$ ($A$~=~Cs, Rb and $B$~=~Cl, Br).\cite{collins1997}

\begin{figure}
\begin{center}
\includegraphics{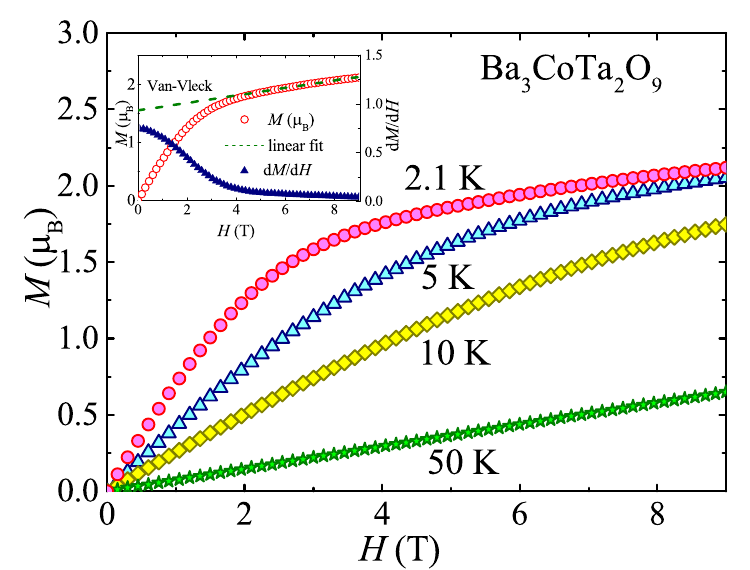}
\caption{Magnetization isotherm ($M$ vs $H$) of Ba$_3$CoTa$_2$O$_9$ measured at different temperatures. Inset: $M$ vs $H$ (left $y-$axis) at 2.1~K showing a nearly linear increase above $H_s$ due to Van-Vleck paramagnetism. $dM/dH$ vs $H$ is also plotted in the right $y-$axis.}
\label{MH}
\end{center}
\end{figure}
In order to obtain further insight into the ground-state properties, we measured $M(H)$ isotherms at different temperatures $T=2.1$~K, 5~K, 10~K, and 50~K, as shown in Fig.~\ref{MH}. At higher temperatures ($T>50~$K), $M$ varies almost linearly with $H$, as expected for AFM materials with a large exchange coupling. For $T<50~$K, it develops a curvature, which is more pronounced at low temperatures. At $T=2.1$~K, the Co$^{2+}$ spins saturate at $H_{\rm s} \simeq 3$~T, above which $M$ increases linearly, but with a much smaller slope.

The inset of Fig.~\ref{MH} shows the magnetic isotherm ($M$ vs $H$) and its derivative $\frac{dM}{dH}$ vs $H$ at $T=2.1~$K in the left and right $y-$axes, respectively. The derivative shows a gradual change of slope with increasing field, and above $\sim 3$~T the change is almost negligible suggesting that the saturation field $H_{\rm s}$ is close to 3~T. This small value of $H_{\rm s}$ is comparable to that reported for Ba$_3$CoNb$_2$O$_9$ \cite{Yokota2014} but much smaller than $H_{\rm s}\simeq32$~T in Ba$_3$CoSb$_2$O$_9$.\cite{Susuki2013} Above $H_{\rm s}$, there is a slow linear increase in $M$, usually attributed to the temperature-independent Van-Vleck paramagnetic contribution, typical for a Co$^{2+}$ ion in an octahedral environment.\cite{Oguchi1965,Lines1963} Similar scenario is also reported for Ba$_3$CoNb$_2$O$_9$ and Ba$_3$CoSb$_2$O$_9$.\cite{Yokota2014,Shirata2012,Susuki2013} A linear fit of the data above $H_{\rm s}$ gives a slope of $\sim 6.4~\times10^{-2}~\mu_{\rm B}/$T, which corresponds to the Van-Vleck susceptibility $\chi_{\rm VV}\simeq3.7\times10^{-2}~$cm$^3$/mol. This is in close agreement with the value $\chi_{\rm VV}\simeq3.4\times10^{-2}~$cm$^3$/mol reported for the TLAF compound Ba$_3$CoNb$_2$O$_9$.\cite{Yokota2014} From the intercept of the linear fit above $H_{\rm s}$ on the $y-$axis, the saturation magnetization was obtained as $M_{\rm s}\simeq1.55~\mu_{\rm B}$. This value of $M_{\rm s}$ ($M_{\rm s}=gJ \mu_{\rm B}$) corresponds to an average $g-$value of $g\simeq3.1$ for $J=1/2$, and the deviation from the free-electron $g$-value of 2.0 is due to the spin-orbit coupling. Such a large value of $g$ is also reported for Ba$_3$CoNb$_2$O$_9$ and Ba$_3$CoSb$_2$O$_9$ \cite{Yokota2014,Susuki2013}. The high and anisotropic $g$-values are expected for Co$^{2+}$ compounds in the octahedral environment due to the large orbital contribution.\cite{Carlin1986} The ESR experiment on the analogous compound Ba$_3$CoSb$_2$O$_9$ indeed reports the high values of $g\simeq3.84$ (for $H\parallel ab$) and $g\simeq3.87$ (for $H\parallel c$).\cite{Susuki2013}

In contrast to Ba$_3$CoNb$_2$O$_9$ and Ba$_3$CoSb$_2$O$_9$, the anomaly in the $M$ vs $H$ curve at $H_{\rm s}\simeq3$~T is not very sharp.\cite{Yokota2014,Susuki2013} The shape of this anomaly depends on various factors such as the $g$-tensor anisotropy. In powder samples, this anisotropy smears the anomaly out. Therefore, the fact that our $M$ vs $H$ curve does not show any sharp anomaly at $H_{\rm s}$ is possibly indicating at an anisotropic $g$-value.

\subsection{Heat capacity}
\label{hc1}
\begin{figure}
\begin{center}
\includegraphics{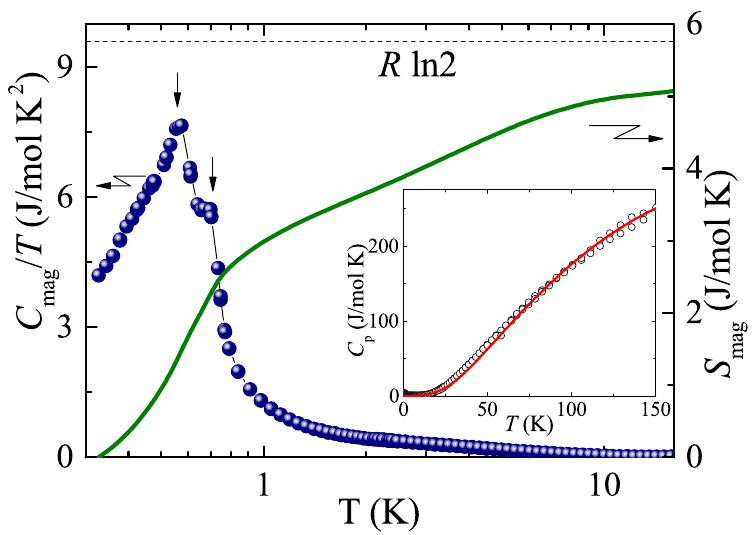}
\caption{$C_{\rm mag}/T$ (left $y$-axis) and $S_{\rm mag}$ (right $y$-axis) as a function of $T$ in the low temperature regime. The downward arrows indicate the two transition anomalies. The horizontal dashed line corresponds to $R\ln 2$. Inset: Temperature-dependent heat capacity $C_{\rm p}(T)$ of Ba$_3$CoTa$_2$O$_9$ measured at zero applied field along with the calculated $C_{\rm ph}(T)$ (solid line).}
\label{HC_Smag}
\end{center}
\end{figure}
Heat capacity $C_{\rm p}(T)$ for magnetic insulators has two major contributions: phononic ($C_{\rm ph}$) and magnetic ($C_{\rm mag}$) parts. At high temperatures, $C_{\rm p}(T)$ is mainly dominated by $C_{\rm ph}$, while at low temperatures it is mostly of magnetic origin. In order to estimate the phonon part of the heat capacity, the $C_{\rm p}(T)$ data at high temperature were fitted by the sum of Debye functions
\begin{equation}\label{hceq}
C_{\rm ph}(T) = 9R\displaystyle\sum\limits_{n=1}^{4} c_{\rm n} \left(\frac{T}{\theta_{\rm Dn}}\right)^3 \int_0^{\frac{\theta_{\rm Dn}}{T}} \frac{x^4e^x}{(e^x-1)^2} dx,
\end{equation}
where $\theta_{\rm Dn}$ are the characteristic Debye temperatures and $c_n$ are the integer coefficients indicating the contributions of different atoms (or group of atoms) to $C_{\rm p}(T)$. Similar procedure has been adopted in various other compounds to estimate the phonon contribution.\cite{Nath2008,Ahmed2015} The inset of Fig.~\ref{HC_Smag} shows $C_{\rm p}$ data as a function of $T$ in zero applied field and the fit (solid line) by Eq.~\eqref{hceq} with $c_1 = 3$, $c_2 = 1$, $c_3 = 2$, and $c_4 = 9$. Here, $c_1$, $c_2$, $c_3$, and $c_4$ represent the number of Ba, Co, Ta, and O atoms per formula unit, respectively. The sum of $c_{\rm n}$ equals to 15, which is the number of atoms per formula unit. Because of the large differences in the atomic masses, we used three different Debye temperatures: $\theta_{\rm D1}$ for Ba$^{3+}$ and Ta$^{3+}$, $\theta_{\rm D2}$ for Co$^{2+}$, and $\theta_{\rm D3}$ for O$^{2-}$. Finally, the high-$T$ fit was extrapolated down to 2~K and $C_{\rm mag}(T)$ was estimated by subtracting $C_{\rm ph}(T)$ from $C_{\rm p}(T)$. $C_{\rm mag}(T)/T$ is plotted as a function of $T$ in Fig.~\ref{HC_Smag}.

To cross-check the reliability of the fitting procedure, we calculated the total magnetic entropy $(S_{\rm mag})$ by integrating $C_{\rm mag}(T)/T$ between 2.1~K and high temperatures as
\begin{equation}
\label{smag}
S_{\rm mag}(T) = \int_{2.1\,\rm K}^{T}\frac{C_{\rm mag}(T')}{T'}dT'.
\end{equation}
The obtained $S_{\rm mag}$ is plotted on the right $y-$axis of Fig.~\ref{HC_Smag}. It reaches the value of $S_{\rm mag} \simeq5.1$~J/mol~K at around 16~K. This value is only slightly smaller than the expected theoretical value [$S_{\rm mag} = R \ln(2J+1)$] of 5.76~J/mol~K for $J=1/2$. This indeed is a strong evidence for the formation of the $J=1/2$ state at low temperatures.

\begin{figure}
\begin{center}
\includegraphics{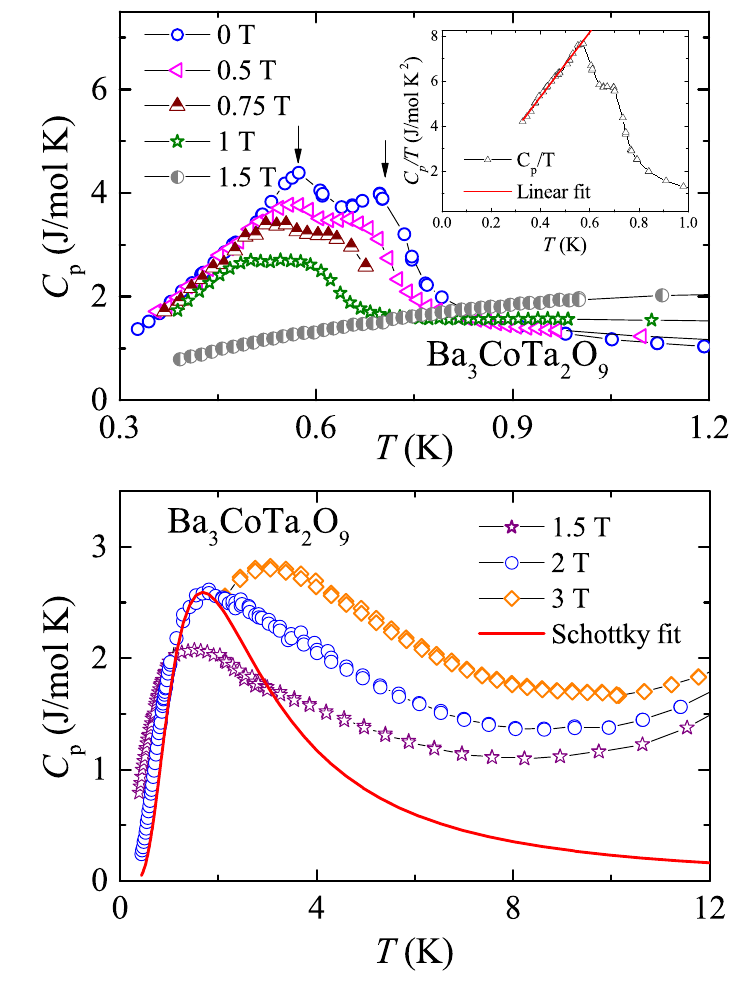}
\caption{Upper panel: $C_{\rm p}(T)$ measured at different applied magnetic fields in the low-temperature region. Inset: $C_{\rm p}/T$ vs $T$ with a linear fit below $T_{N2}$. Lower panel: $C_{\rm p}(T)$ measured at higher applied fields. The solid line is the simulated curve using Schottky expression Eq.~\eqref{sch}.}
\label{HC_diffH}
\end{center}
\end{figure}
The $C_{\rm p}(T)$ measured at different applied fields from 0~T to 1.5~T in the low temperature regime is shown in the upper panel of Fig.~\ref{HC_diffH}. In zero field, it exhibits two anomalies suggesting two successive magnetic transitions at $T_{N1}\simeq0.70$~K and $T_{N2}\simeq0.57$~K. Below $T_{N2}$, the $C_{\rm p}(T)$ data were fitted using a power law $C_{\rm p}\propto T^{\alpha}$ with the exponent $\alpha\simeq2.05$ (inset of the upper panel of Fig.~\ref{HC_diffH}). A quadratic temperature dependence of $C_{\rm p}$ in the ordered state is expected for 2D lattices.\cite{Nakatsuji2005} With increasing field, the transition peaks get smeared and shift towards lower temperatures before they completely disappear at $H\geq1.5~$T. With increasing field, a broad feature appears at about $T=2$~K for $H=1.5$~T, which can be identified as the Schottky anomaly arising due to the Zeeman splitting of the $J=1/2$ energy levels at low temperatures (lower panel of Fig.~\ref{HC_diffH}). As the field increases, the Schottky anomaly is observed to shift towards higher temperatures. It is to be noted that, although the peak positions and their heights at the magnetic transition as well as Schottky anomaly regimes are changing with magnetic field, the value of $S_{\rm mag}$ at 16~K remains close to 5.76~J/mol~K, irrespective of the magnitude of the applied field. For $H=2$~T, a theoretical curve was simulated using the Schottky expression,
\begin{equation}
\label{sch}
C_{\rm Sch}=N_A k_B\left(\frac{g\mu_{\rm B} H}{2k_B T}\right)^2 \frac{1}{{\rm cosh}^2(g \mu_{\rm B} H/2k_B T)}.
\end{equation}
This expression represents the Schottky specific heat due to the Zeeman splitting of the $J=1/2$ state without exchange interactions between the spins. As shown in Fig.~\ref{HC_diffH}, the low temperature data match well with the simulated curve with $g\simeq3.1$, further confirming the large value of $g$ inferred from the magnetization measurements. The deviations between the simulated curve and experimental data at higher temperatures are attributed to the phonon contribution of the heat capacity.

\section{Discussion}
\begin{figure}
\begin{center}
\includegraphics{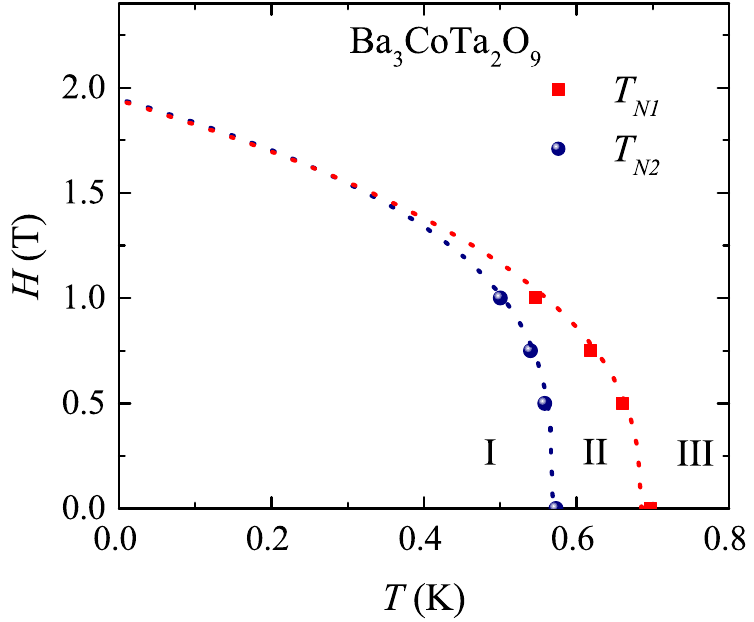}
\caption{The $H-T$ phase diagram for Ba$_3$CoTa$_2$O$_9$ showing three distinct phase regimes. The dotted lines represent the power law behavior.}
\label{PD}
\end{center}
\end{figure}
%The large and positive value of $\theta_{CW}$ ($\simeq 50.1$~K) indicates dominant AFM interaction in the compound. Strong magnetic frustration is expected due to triangular geometry and the AFM interactions in Ba$_3$CoTa$_2$O$_9$. In order to quantify the magnetic frustration, the frustration parameter $f=|\theta_{\rm CW}|/T_{N1}$ was calculated to be $f\simeq71.9$.\cite{Ramirez1994} Such a large value of $f$ clearly suggests the strongly frustrated nature of Ba$_3$CoTa$_2$O$_9$. Assuming that each Co$^{2+}$ spin interacts only via exchange interaction $J$ with the $z=6$ nearest neighbor spins on the triangular layer, the Hamiltonian can be written as $H=J\sum_{i,j}S_i.S_j$. According to mean-field theory, this leads to a CW temperature $\theta_{\rm CW}=zJS(S+1)/3k_B$. For Ba$_3$CoTa$_2$O$_9$, taking $S=3/2$, our experimental value of $\theta_{\rm CW}$ gives $J/k_B\simeq6.68~$K.

Ba$_3$CoTa$_2$O$_9$ shows two successive magnetic phase transitions at low temperatures. It has been theoretically predicted that double magnetic transitions can occur in TLAFs when the magnetic anisotropy is of the easy-axis type, while a single transition is expected for easy-plane type anisotropy.\cite{Matsubara1982,Miyashita1985} In TLAFs with the easy-axis anisotropy, the 120$^\circ$ state is often preceded by a collinear state. Our zero-field measurements show two transitions that can be thus associated with the collinear state below $T_{N1}$ and the 120$^{\circ}$ state below $T_{N2}<T_{N1}$. The putative easy-axis anisotropy in Ba$_3$CoTa$_2$O$_9$ is similar to that reported for other TLAFs, Ba$_3$CoNb$_2$O$_9$, RbMn(MoO$_4$)$_3$, CsMnI$_3$, and CsNiCl$_3$.\cite{Yokota2014,Ishii2011,Harrison1991,Kadowaki1990,Clark1972,Kadowaki1987} The temperature range of the intermediate phase can be used to obtain ($T_{N1}-T_{N2}$)/$T_{N1}$ that roughly quantifies the size of the easy-axis anisotropy with respect to the intra-layer exchange coupling. The narrow intermediate phase ($\sim0.18$) seen in Ba$_3$CoTa$_2$O$_9$ suggests that the easy-axis anisotropy is significantly smaller than the intra-layer coupling.

The $H-T$ phase diagram derived from the field-dependent $C_{\rm p}(T)$ measurements is shown in Fig.~\ref{PD}. It exhibits three distinct phases denoted by I, II, and III. Region III corresponds to the paramagnetic phase, while regions II and I are expected to be the collinear and 120$^\circ$ spin states, respectively, in line with the theoretical predictions and the subsequent experimental realizations in Ba$_3$CoNb$_2$O$_9$ and RbMn(MoO$_4$)$_3$.\cite{Yokota2014,Ishii2011,Chubukov1991,Kawamura1985} %The obtained phase diagram is comparable to that of other cobaltate compounds Li$_2$CoW$_2$O$_8$ and Pb$_3$TeCo$_3$V$_2$O$_{14}$ \cite{Muthuselvam2014,Markina2014}, both of which are TLAFs with weak inter-layer coupling and successive AFM transitions, although a canted AFM phase appears in both of them at higher fields. It is to be noted that the decreasing range of field for which $uud$ phase is stabilized, is similar to the case for classical spins, where magnetic field range of $uud$ phase decreases with decreasing temperatures and vanishes at zero temperature.\cite{Chubukov1991,Miyashita1986}

The data points in Fig.~\ref{PD} are fitted by a power law,
\begin{equation}
H=H_c\left(1-\frac{T}{T_N}\right)^\beta,
\label{ce}
\end{equation}
where the critical field $H_c$, $T_N$, and critical exponent $\beta$ are the fitting parameters. The $\beta$ value reflects the universality class and dimensionality of the spin system. The values obtained for $T_N$ are consistent with the transition temperatures at zero applied field and the value of $\beta\simeq0.31$ (for $T_{N2}$) and 0.38 (for $T_{N1}$) are close to the value of $\beta=1/3$ predicted by the mean-field theory for three dimensional (3D) spin systems.\cite{Nath2009}

\section{Conclusion}
Magnetic properties of the triangular antiferromagnet Ba$_3$CoTa$_2$O$_9$ were investigated. At low temperatures, it undergoes two successive AFM transitions at $T_{N1}\simeq0.70~$K and $T_{N2}\simeq0.57~$K with a narrow intermediate phase, which is due to weak easy-axis magnetic anisotropy. Co$^{2+}$ adopts the $J=1/2$ state at low temperatures, whereas at higher temperatures the $S=3/2$ state is observed. The magnetization saturates already at $H_{\rm s}\simeq3~$T suggesting weak magnetic interactions in the system. The $H-T$ phase diagram deduced from the heat capacity measurements shows three distinct phases. The above peculiar features render Ba$_3$CoTa$_2$O$_9$ a model TLAF compound for further experimental and theoretical studies.

\acknowledgments
We thank Alexander Tsirlin for fruitful discussions.

%\bibliographystyle{prsty}
%\bibliography{reff_BCTO}

\end{document}